# A High-Level Modeling Language for the Efficient Design, Implementation, and Testing of Android Applications

**John Abou-Jaoudeh[1], Kinan Dak-Al-Bab[1], Mostafa El-Katerji[1], Yliès Falcone[2], Mohamad Jaber[1]**

[1] American University of Beirut, Beirut, Lebanon
   e-mail: {jia03,kmd14,mme85}@mail.aub.edu, mj54@aub.edu.lb
[2] Laboratoire d'Informatique de Grenoble, Université Grenoble-Alpes, Grenoble, France
   e-mail: Ylies.Falcone@ujf-grenoble.fr

**Abstract.** Developing mobile applications remains difficult, time consuming, and error-prone, in spite of the number of existing platforms and tools. In this paper, we define MoDroid, a high-level modeling language to ease the development of Android applications. MoDroid allows developing models representing the core of applications. MoDroid provides Android programmers with the following advantages: (1) Models are built using high-level primitives that abstract away several implementation details; (2) It allows the definition of interfaces between models to automatically compose them; (3) Java native android can be automatically generated along with the required permissions; (4) It supports efficient model-based testing that operates on models. MoDroid is fully implemented and was used to develop several non-trivial Android applications.

## 1 Introduction

Android is the most popular platform for mobile devices, with over 84% of market share at the end of 2014. Yet, creating a correct and efficient Android application remains a difficult endeavor for several reasons that can be categorized under design or testing issues.

*Issues when designing an Android application.* First, the programming model in Android involves different components (e.g., `Activity`, `Service`, `BroadcastReceiver`, `ContentProvider`, etc.), with a complex interaction model between these components (e.g., `Handler`, `Intent`, etc.). Second, to separate the internal representation of information from its presentation to the user, most of the frameworks supporting the development process use the Model-View-Controller (MVC) design pattern to split an application into three interconnected parts. However, as applications become more complex, the MVC pattern must be augmented with a new paradigm that guides developers on how to split the core of an application into different interconnected parts. Such paradigm shall facilitate and encourage the concurrent development of an application by several developers. Third, Android provides a protection mechanism to devise-specific features (e.g., GPS, camera, vibrator, internet, SMS, address book, SD card, etc.) by offering a specific set of programmatic APIs to access them. Then, the application configuration file (`AndroidManifest.xml`) must explicitly include access permissions for all features that are used within the application. At installation, the application is given permission to the corresponding features (from the configuration file) and the user will be aware about the required permissions. If an application calls an API to access a specific feature that requires a permission access and the configuration file does not contain that access permission, a runtime exceptions will be raised at the start-up of the application. Clearly, users prefer applications with minimum set of permissions. This protection mechanism is often error-prone and in most of the cases developers end up using permissions they do not require in their code, or the opposite [4].

*Issues when testing an Android application.* On the other hand, ensuring that applications are performing as required has become more challenging given the daily dynamic change in the domain of mobile technology. Application users mainly face problems of the following kind: incorrect behavior, crashes, and Application becoming Not Responsive (ANR), etc. Keeping in mind the complexity of mobile application development, and the inability to eliminate bugs and errors, an essential component of mobile development is testing. The process of Mobile Application Testing is used to detect the errors that might have occurred during the development of the application, to ensure that user expectations are met, and to make sure that applications have been executed properly. This is essential to be done by application developers who aim to keep their customers satisfied, and entertained by the final product.

*Contributions.* The challenges of programming mobile applications have prompted us to reconsider the best practices of their design development. For this purpose, a framework with the following features is desirable: (1) the framework should abstract away different implementation details; (2) decompose the development process into different stages; and (3) include automated code manipulation and generation. To do so, we define a Meta-Model for the development of mobile android applications. Meta-modeling drastically improves flexibility of development, hence allows us to manage applications more easily.

The Meta-Model consists of a set of modules that represent Graphical User Interfaces (GUIs) and their respective handlers in an abstract and a simpler way than Native Java Android. We implement the Meta-Model along with several modules in MoDroid to tackle the aforementioned problems. MoDroid contains the following modules:

1. A composition module takes as input Android Java models and the connections between them. The composition module allows to easily parallelize the development process.
2. A permission analysis automatically discovers the required permissions of an application.
3. A code generator automatically generates native Android Java code given an android Java model.
4. An activity-builder module automatically builds an activity in the Android Java model given an `XML` file representing that activity.
5. An efficient model-based testing that allows to easily write test cases using high-level primitives and to efficiently execute them.

Our framework facilitates and speeds-up the development process. It transforms an Android application into an Android Java model that is compliant to the Meta-Model and contains all the necessary information about the application. The current version of our Meta-Model covers a subset of Android API that includes all the main constructs and functionalities. Consequently, it is designed with backward compatibility in mind so that developers can write native Android code within the model to use features currently not covered by the Meta-Model.

*Paper organization.* The rest of this paper is structured as follows. Section 2 presents the Meta-Model. The following sections present the components associated to the Meta-Model: model composition is presented in Sec. 3; and automatic permissions detection is presented in Sec. 4; model-based testing framework is presented in Sec. 5; and automatic code generation (from high-level model to native android) is presented in Sec. 6. Sections 7 and 8 describes MoDroid, a full implementation of our framework and some benchmarks. Section 9 discusses related work. Section 10 draws some conclusions and perspectives.

## 2 The Android Meta-Model

The Meta-Model consists of a set of modules used to model the core of an Android application. The Meta-Model allows to model an Android application as a Java object. The modeling process abstracts away implementation details. Moreover, the resulting object model can be easily and efficiently manipulated by applying model transformation and composition as described in the remainder of this paper.

The Meta-Model consists of a hierarchy of classes. The top element of the hierarchy is the project: `LibModel`. Each instance of this type represents an independent application. A `LibModel` consists of a set of activities mapped to names, global variables, and meta-information related to the project.

An activity `LibActivity` is the android equivalent of a window or frame. The developer can create instances of `LibActivity`, fill it up with GUI elements, and then add it to a `LibModel`. A `LibActivity` can contain GUI elements (e.g., layout, button, etc.), packaging information, and *activity scope variables*. The developer can also provide methods for handling events related to the activity's life cycle: `onCreate`, `onStop`, etc. Moreover, `LibActivity` has a constructor that takes an XML file as argument containing a view description of the activity and automatically instantiates the corresponding object. That is, we can still benefit from MVC design pattern supported for native android development.

GUI elements, also called views, are the building blocks of an application. All GUI elements inherit their basic attributes from `LibView`, an abstract class that contains the basic attributes and methods for the manipulation of appearance of an element such as width, height, padding, etc. Views are categorized into Controls, and Layouts. A view can be either added to an Activity or to a layout. The controls currently provided by the Meta-Model, prefixed with `Lib`, are the following: `Button`, `ImageButton`, `TextView` (equivalent to a Label), `TextField`, `ToggleButton` (on/off button), `Spinner` (similar to drop-down list), `RadioButton`, `CheckBox`, etc.

Layouts are special views that can contain other views. They control the position of the view within the activity. A `layout` is treated as a View. It has its own attributes such as width, height, and others. It can be added to activities, or to other layouts. The layouts provided by the Meta-Model, prefixed with `Lib`, are the following: `LinearLayout` (views are placed in order in a line; can be horizontal, or vertical),

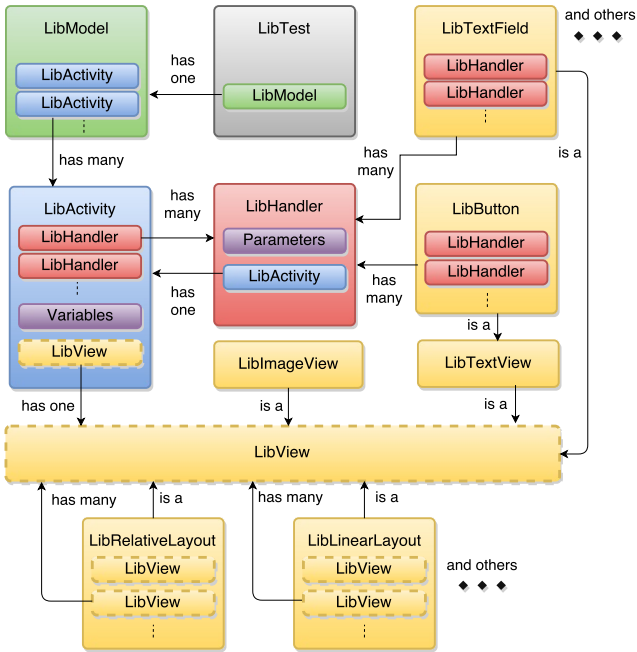

**Fig. 1.** Basic elements of the Meta-Model

`RadioGroup` (a LinearLayout that acts as a RadioButton group as well), `FrameLayout` (displays all views in the same position above each other), `RelativeLayout` (controls the position of views by using them as anchors), and `TableLayout` (organizes the views into rows and columns).

These views cover all the basic elements of Android applications. Moreover, it is possible to extend the Meta-Model by adding more views in an easy and modular way. Figure 1 depicts the basic elements of the Meta-Model. Hereafter, we show a step-by-step how to build a simple health application using our paradigm. The health application consists of two basic modules: (1) Body Mass Index (BMI); and (2) Menu Planner/Meal Planner. The BMI module is composed of two activities. The first activity manages user inputs (weight and height) and computes the BMI. Then, it sends the computed value to the second activity. If the user does not enter a value and clicks on compute, the phone vibrates signaling an error. Moreover, the user inputs are stored in the activity scope variables. The second activity is where the BMI value is displayed. From this activity, a user may either navigate back to activity one or navigate to Menu Planner/Meal Planner module. Listing 1 shows a snapshot of the code of BMI module.

### 2.1 Handlers

Some views have special events that trigger specific handlers (e.g., on button click). A developer can either write a method which handles the event or use some pre-defined shortcuts. The code within the handlers can use functionalities of the Meta-Model or can directly use native Android code. Views can be accessed within handlers by passing them as parameters of the handler method. A handler can be used for the communication between activities. For example, when a button is clicked or some text filed gets modified, one common functionality is to go to another activity. For a given view, one specifies its handler method by calling `setOnClickHandler`. The Meta-Model simplifies control transfer by using high-level shortcut. For instance, within a handler, `startActivity` method redirects to another activity by taking the name of the activity and any view objects as parameters. Another shortcut is to directly specify the next activity in the `setOnClickHandler`.

Data parameters can be sent with a control transfer to communicate between activities. These parameters can be passed either as parameters (1) to `startActivity` along with the next activity; or (2) directly to `setOnClickHandler`.

Listing 2 shows the code of the button from the first activity where its handler computes the BMI value and send it to the second activity. Note that, if the user does not enter a value and clicks on compute, the phone vibrates signaling an error.

These parameters can be accessed in the main method by using a special formatted string (`@param_{i}` to get the $i^{\text{th}}$ parameter). Within a handler, these parameters can be also accessed by calling `LibActivity.getParameter(i)` to get the $i^{\text{th}}$ parameter. Listing 3 shows a snapshot of the code that sets some of the views of the second activity. It sets the the value of a text view to the passed parameter that comes from the first activity. Also, it uses a shortcut to set the handler of the button that redirects to the first activity.

### 2.2 Resource Management

One of the most effort consuming task in developing Android applications is resource management: images, application icons, and other types of resources. These resources are copied to specific folders within the resource folder. In our Meta-Model, resources are automatically added and generated into their corresponding folders. For example, to use an image, the developer only needs to add the path of the image/icon to be used. Listing 4 shows an example that specifies the icon of an application, displays an image, and create a buttong with an image displayed.

## 3 Projects Composition

Decomposing projects into smaller parts is a key concept in software engineering. Using the Meta-Model, it is possible to develop several models and automatically compose them according to a user-provided configuration.

**Listing 1.** Snapshot of the code of BMI module.

```
1  LibModel bmiModel = new LibModel("bmiModel", "health.app", "John");
2  LibActivity userInputActivity = new LibActivity();
3  LibActivity resultActivity = new LibActivity();
4  bmiModel.addActivity(userInputActivity, "userInputActivity");
5  bmiModel.addActivity(resultActivity, "resultActivity");
6  setUserInputActivityLayout( userInputActivityLayout );
7  setResultActivityLayout( resultActivityLayout );
8  ...
```

**Listing 2.** Example of a handler with data transfer.

```
1  calculateButton.setOnClickHandler("Handler:health.BMI.calculate", height, weight);
2
3  // package health.BMI
4  public void calculate(LibView ht, LibView wt) {
5     if(!ht.getText().equals("") && !wt.getText().equals("")) {
6        double val = computeBMI (ht, wt);
7        LibModel.startActivity("resultActivity", val);
8     }
9     else {
10       Vibrator v = (Vibrator) getSystemService(Context.VIBRATOR_SERVICE);
11       if(v.hasVibrator()) v.vibrate(500);
12    }
13 }
```

**Listing 3.** Example of shortcut handler and data access.

```
1  bmiValueText.setText("@param_0");
2  ...
3  goBackButton.setOnClickHandler("GoToActivity:userInputActivity");
```

The composition operation takes as input a configuration file that specifies the links between the interfaces of models. Each link specifies some control and data transfer that have to occur upon the occurrence of an event in the models: the activity from another project that has to be executed and the parameters that have to be sent.

*Principles.* Given $n$ models $m_1, m_2, \ldots, m_n$, where $m_i$ consists of $a_1^i, a_2^i, \ldots, a_{I_i}^i$ activities. Recall that each activity has views that may have handlers. Each handler runs some code that may transfer the control to another activity that can be an identified activity in the model or a symbolic activity (i.e., an activity which is identified by a symbolic value). Symbolic activities within a handler are specified by using method `goToUnknown` that takes an identifier and a set of objects (to be passed to the other activity) as parameters. A model that has a handler that transfers to a symbolic activity is considered as a *partial model*.

If a handler only redirects to a symbolic activity, it is possible to use pre-defined high-level shortcut to do so. At an abstract level, the composition module relies on two functions: interface that returns the symbolic activities in a model, and, configuration that associates (concrete) activities to symbolic activities. The definition of function interface is obtained by an automatic analysis of models (see Sec. 7). Function configuration is defined by the user through a configuration file. A configuration file is of the form depicted in Listing 5. It first contains the new project name, package, author and main activity. Then, it defines the mapping between identifiers and activities of different models.

Let $m_i$ be a partial model with some of its handlers associated to symbolic activities $id_1^i, id_2^i$ (interface$(m_i) = \{id_1^i, id_2^i\}$). Let $a_k^j$ be an activity of model $m_j$, one can have configuration$(id_1^i) = a_k^j$, which means that identifier $id_1^i$ of model $m_i$ is mapped to activity $a_j^k$ of model $m_j$.

*Example.* Figure 2 is an example of two partial models `M1` and `M2`. The handler of button `button2`, a handler of activity `A2` and the handler of button `button3` redirect to symbolic activities though interfaces `I1`, `I2` and `I3`,

**Listing 4.** Example of resource management.

```
1   ...
2   model.setIcon("images/application.jpg"); // sets the application icon
3   // Create a label to display the given image.
4   LibImageView imageView = new LibImageView("images/image.jpg", ...);
5
6   // Create a button with an image displayed on it.
7   LibImageButton imageButton = new LibImageButton("images/button.jpg", ...);
```

**Listing 5.** General shape of a configuration file.

```
1   <New Project Name>
2   <New Project Package>
3   <New Project Author>
4   <Model>.<Activity>; //indicates the main activity of the composed project
5   <Model>.<Unknown ID> -> <Model>.<Activity Name>; // mapping
6   <Model>.<Unknown ID> -> <Model>.<Activity Name>; // mapping
7   <Model>.<Unknown ID> -> <Model>.<Activity Name>; // mapping
8   ...
```

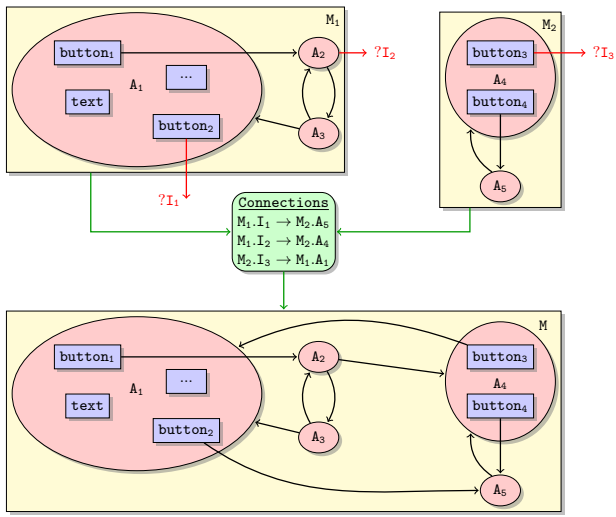

**Fig. 2.** Example of models composition.

respectively. The configuration file connects $I_1$, $I_2$ and $I_3$ to activities $A_5$, $A_4$ and $A_1$, respectively.

Listing 6 shows a snapshot of the shortcut handler of the button from the second activity (result activity) of the health application that redirects to a symbolic activity of a different model through the interface `menuPlannerInterface`.

Finally, models can be composed to build the final project by using `LibModel`'s constructors that takes a configuration file and a set of models. The composition of BMI and Menu Planner modules is depicted in Listing 7.

Listing 8 shows the configuration file that connects (1) the menu planner interface of BMI calculator model to user information activity of the menu planner model;

and (2) the BMI calculator interface of menu planner model to user input activity of BMI calculator model.

Note that, a set of models can be composed successively to build the final model. Listing 9 shows an example of successively composing three models.

Although mobile applications almost certainly harbors undetected errors, using models composition approach, it is possible to directly apply software testing paradigm to reduce and locate them: unit and integration testing. This can be done by testing partial models separately (unit testing) to find local errors and then test the complete model (integration testing) to find interface errors.

## 4 Permission Auto-detection and Generation

Manually managing permissions in the configuration file is time consuming. It often entails several compilation attempts of the application to narrow the proper set of required permissions. Consequently, most of the developers add permissions more than it is needed which contradicts with the users' preferences. For example, to use the phone's vibrator, one needs to retrieve the vibrator object using the method `getSystemService(Context.VIBRATOR_SERVICE)`, then call one of the following methods: `hasVibrator()`, `vibrate()`, or `cancel()`. Note that, method `hasVibrator()` returns a boolean and does not require the vibrate permission (`android.permission.vibrate`), while `cancel()` and `vibrate()` do. Listing10 shows an example of native Android Java code that calls `hasVibrator()` but does not require permission access which is actually not needed. Intuitively, developers may

**Listing 6.** Example of unknown shortcut handler.

```
1  menuPlannerButton.setOnClickHandler("GoToActivity:Unknowns(menuPlannerInterface)");
```

**Listing 7.** Composition of BMI and Menu Planner modules.

```
1  LibModel healthAppModel = new LibModel("config.txt", bmiCalculatorModel, menuPlannerModel);
```

**Listing 8.** Configuration file connecting BMI and Menu Planner models.

```
1  Health App // project name
2  health.app // project package
3  John // project author
4  bmiCalculatorModel.userInputActivity // main activity of the composed model
5  // connections/mapping
6  bmiCalculatorModel.menuPlannerInterface -> menuPlannerModel.userInformationActivity
7  menuPlannerModel.bmiCalculatorInterface -> bmiCalculatorModel.userInputActivity
```

assume that method `hasVibrator()`, or/and class method `getSystemService()` requires permission `android.permission.VIBRATE` and adds it to the manifest configuration file. Note that, if one replaces line 8 with `v.vibrate(500)`, the permission access would be required only for mobiles that have a vibrator. Consequently, code modifications require a manual reconsideration of the required permissions. In our

**Listing 10.** Example of native Android Java code that does not require permission.

```
1  @Override
2  protected void onCreate(Bundle savedInstanceState) {
3      super.onCreate(savedInstanceState);
4      setContentView(R.layout.activity_main);
5
6      Vibrator v = (Vibrator)
7          getSystemService(Context.VIBRATOR_SERVICE);
8
9      if(v.hasVibrator()) {
10         Toast.makeText(this, text, duration).show();
11     }
12 }
```

case, APIs to access devise-specific features are called within handlers of listener GUI elements. Note that some external libraries may call some of these APIs. Our permission detection/generation module must take into account: (1) modification (add/remove/update) of permissions; (2) modification (add/remove/update) of APIs; (3) modification (add/remove/update) of external library that may call those APIs. In other words, any of these modifications should not drastically affect the code that automatically detects and generates permissions.

For this, we define a set of templates that represent all the APIs that requires a permission. For instance, object initializations (constructors), method calls (method name, parameter types, calling object's type), etc. This gives us maintainability for future permission modification as well as ease to extend our supported set of permissions. We define two types of templates `permissions.xml` and `permissionExternals.xml` that contain templates for native APIs and external library APIs, respectively, that require permission access. The template file is of the form depicted in Listing 11. The template depicted in Listing 11 defines all the API calls shown in 12 that require permission `PERMISSION_1`:

For example, the template for permission `android.permissions.VIBRATE` is depicted in Listing 13. From the template of permission `android.permissions.VIBRATE`, we can deduce that permission `android.permissions.VIBRATE` is required whenever one of the lines of code in Listing 14 is detected.

## 5  Model-based Testing

In order to integrate efficient model-based testing in our framework, we extend our model to be executable. That is, each model can be represented as a state consisting of the current activity, the value of the views, the value of the activities scope variables and global variables. We implement all the functionalities to perform operations on a given model. For example: (1) modify or get the value of a view; (2) perform click/event. In order to perform a click, we use Java reflection to execute the handler of a corresponding view (e.g., button). Performing operations modify the state of the model accordingly.

**Listing 9.** Successive composition of models.

```
1  LibModel model12 = new LibModel("config1.txt", model1, model2);
2  LibModel model123 = new LibModel("config2.txt", model12, model3);
```

**Listing 11.** General shape of a template file for a given permission.

```xml
<permission name="PERMISSION_1">
    <class name="Class_1">
        <method name="method_1">
            <parameters>
                </parameters>
        </method>

        <method name="method_2">
            <parameters>
                </parameters>
        </method>
        <method name="method_3">
            <parameters>
                </parameters>
            <parameters>
                </parameters>
        </method>
    </class>

    <class name="Class_2">
        <method name="method_4" />
        <method name="method_5" />
    </class>

    <class name="Class_3" />

</permission>
```

**Listing 12.** API calls requiring permission PERMISSION_1.

```
2   (Class_1).method_1(param1);

4   (Class_1).method_2();

6   (Class_1).method_3(param2, param3);

8   (Class_1).method_3(param4);

10  (Class_2).method_4(...);

12  (Class_2).method_5(...);

14  Class_3 var = new Class_3(...);
```

**Listing 13.** Template for permission android.permissions.VIBRATE.

```xml
<permission name="android.permissions.VIBRATE">
    <class name="Vibrator">
        <method name="vibrate" />
        <method name="cancel">
            <parameters>
                </parameters>
        </method>
    </class>
</permission>
```

**Listing 14.** API calls requiring permission android.permissions.VIBRATE.

```
1   // v is an object of type Vibrator
2   // E.g., Vibrator v = (Vibrator)
3   //    getSystemService(Context.VIBRATOR_SERVICE);
4
5
6   // vibrate(long milliseconds) method
7   v.vibrate(500);
8
9   // vibrate(long[] pattern, int repeat) method
10  v.vibrate({{12}, {23}, {12}}, 50);
11
12  // cancel() method
13  v.cancel();
```

The model-based testing framework consists of a module **LibTest** that allows to perform high-level operations on the model under test (e.g., setText, click, etc.). **LibTest** takes a model under testing as input with an optional entry point (i.e., name of an activity) and a set of test cases to be performed.

Recall that, it is possible to test partial models separately (unit testing) to find local errors and then test the composed model (integration testing) to find interface errors.

Listing 15 shows an example of some test cases of BMI calculator model. It mainly tests the redirection of activities and the computation of BMI. It consists of the following steps:

1. Create a **LibTest** instance that takes the model as input. Note that, it is possible to give an activity entry point of the model.
2. Set the weight and the height values and check if the values have been set properly.

3. Perform click on `calculateButton` button and check if (1) the next activity is the result activity; and (2) the BMI was correctly computed.

Note that, if one performs an operation on a view that does not exist in the current activity, an exception is thrown.

## 6 Code Generation

Finally, given an Android model we implement a module that generates equivalent native Android code (along with its resources, manifest configuration file, etc.). This is done by calling `generate(path)` method on a given model. The generated code preserves the order of statements and comments. This allows to easily integrate other functionalities to the generated code.

An example of code generation is depicted in Listing 16.

Listing 16. Example of code generation.

```
1  public class Application {
2     public static void main(String[] args) ... {
3        ...
4        healthAppModel.generate("gen/");
5     }
6  }
```

### 6.1 Cloud-based Compilation

Android SDK (Software Development Kit) is a set of components that include libraries, a debugger, a handset emulator, and others. Its main role in development is to generate native Android application executable files (`.apk`). Android SDK is a heavy module that requires memory, and time.

For this, we have developed a web service, and placed it online to generate an application's executable without installing the SDK. We have configured a server on the cloud with: (1) all the updated Android SDK libraries; (2) `ant-apache` which is a command based tool to create, and update an application given its source code; (3) compiled version of MoDroid. The web service takes as parameter a model of an Android application developed using MoDroid. The server compiles an application and generates an executable file (`.apk`) ready to be installed on Android devices, and shared on Google Play Store. As a plus, in order to efficiently test an application on different real devices, the web service, can send the generated executable to a list of email addresses (application beta testers).

Figure 3 illustrates the process of uploading a MoDroid Android Java code, and receiving an executable native Android code compiled by our server.

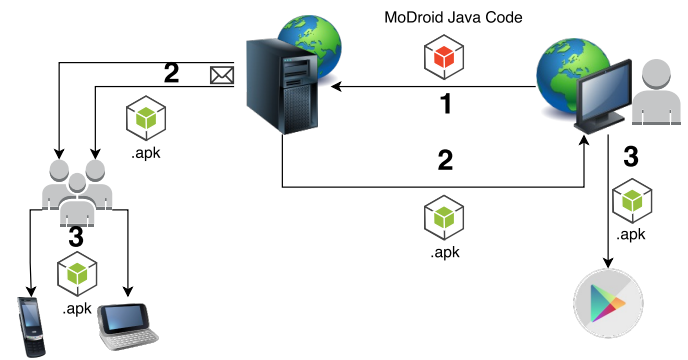

**Fig. 3.** Cloud-based Compilation

## 7 Tool-set - MoDroid

MoDroid[1] implements the Meta-Model and its supported tools: models composition, permission detection, testing and code generation. The tool is packed and compiled into a single `jar` file. The jar file must be imported as a library to the project being developed.

To promote extensibility and modularity of MoDroid we implement a visitor pattern that traverses the tree structure (GUI element, handlers, etc.) of an Android model. The pattern takes as input an interface that declares methods to be executed depending on the node that was localized. We have developed several implementation of that interface:

1. Implementation to detect unknown interfaces (symbolic activities) used in models composition.
2. Implementation that takes templates representing all the APIs that require permissions and detect the required permissions accordingly.

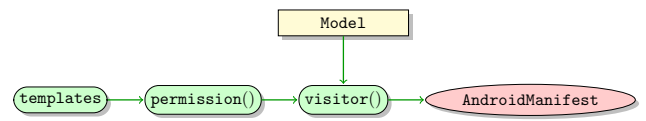

3. Implementation to make the model executable by performing operations on a view (e.g., `LibText`) that are used by model-based testing module.

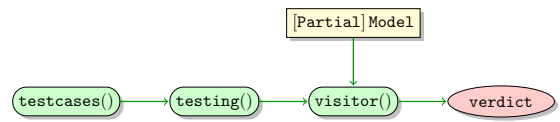

4. Implementation to generate equivalent native Android code (along with its resources, manifest configuration file, etc.) from an Android model. Code generation module uses `antlr` and template engine library `StringTemplate` [14] for parsing handlers and generating native Java Android from an Android model.

---
[1] `http://ujf-aub.bitbucket.org/modroid/`

**Listing 15.** Example of test cases.

```java
@Test
public void testcase1(LibModel bmiModel) {
    try {
        LibTest test = new LibTest(bmiModel);
        test.setText("height", "175");
        assertEquals("Incorrect Height", "175", test.getText("height"));

        test.setText("weight", "70");
        assertEquals("Incorrect Weight", "70", test.getText("weight"));

        test.click("calculateButton");

        assertEquals("Incorrect Activity", "resultActivity", test.getCurrentActivityName());
        assertEquals("Incorrect Value", "22.9", test.getText("value"));
    } catch (ElementNotFoundException e) {
        fail("Element Not Found: " + e);
    }
}
```

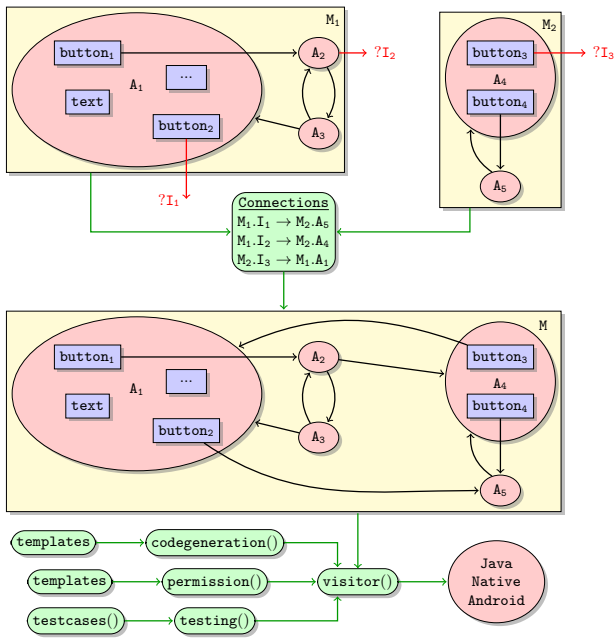

**Fig. 4.** Development design-flow in MoDroid.

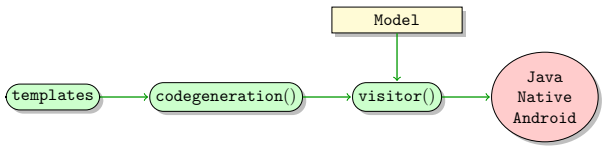

the Meta-Model, one can still benefit from MVC design pattern supported for native Android development.

Second, given a configuration file describing the mapping between models, we generate and test the final model of the application. It is worth mentioning that, we can build several applications given different mappings without any modifications of the models. Finally, a native Java Android code is generated including all the permissions that are required by the core of the application. Developers may edit the generated code to add any extra functionalities.

## 8 Experimental Results

We have developed several applications using both native Java Android and MoDroid. Both versions of the code have the same design and perform exactly the same functions. Table 1 compares the number of lines of code between native Java android, MoDroid, and automatically generated code.

It is clear that building an Android model drastically reduces the number of lines of code. Moreover, it is much less time consuming w.r.t. writing native Java Android. We notice an overhead of ca. 25% in the automatically generated code. This overhead is mainly due to the code generation of handlers. In fact we duplicate handlers of different views which can be technically eliminated by creating only one method for the same handler code of different views.

Moreover, we have conducted other benchmarks to compare the performance of our model-based testing framework and the following tools that are currently widely used: Robolectric, Robotium, and Espresso on both an Emulator and a real device.

Figure 4 shows the development design-flow which is based on MoDroid.

First, models are built and tested separately using high-level primitives provided by MoDroid. Recall that, it is also possible to build models without their handlers (e.g., only GUI layouts) from an XML file and then handlers can be programmatically integrated. That is, using

| Application Name | MoDroid | Generated | Native | Written Code Reduction | Overhead Code Generated |
|---|---|---|---|---|---|
| Breadcrumb Viewer | 63 | 329 | 276 | 77% | 17% |
| Guessing Game | 158 | 340 | 246 | 35% | 27% |
| Scientific Calculator | 180 | 377 | 282 | 36% | 25% |
| Volleyball Statistics | 137 | 702 | 510 | 73% | 27% |

**Table 1.** Code comparison.

Robotium and Espresso perform actions on an emulator or on a real device; whereas Robolectric and MoDroid testing do not need an emulator nor a real device. Taking this factor into consideration, we would expect our testing framework and Robolectric to have a better performance.

The first benchmark was performed on the scientific calculator application that we developed using MoDroid. The test actions were simply to click on values and operations; then to check the output of the calculator.

Table 2 shows a comparison of the time taken to perform test cases that require 10, 25, 50 up to 1 million operations by all the tools. Operations consist of performing clicks and text value modifications and searches.

As expected, Robolectric and MoDroid drastically outperform Robotium and Espresso. The results were close between Robolectric and MoDroid if we take into account the initialization phase required by Robolectric. The time taken to perform test cases requiring one million operations with Robolectric is 27 seconds as opposed to 7.7 seconds using MoDroid.

The second benchmark was performed on a Volleyball Statistics application developed using MoDroid. It is composed of two activities. The first activity is the splash screen which contains a button to navigate to the second activity where statistics are done. The second Activity is composed of two teams and the players for each team. Each player has two buttons to increment and decrement the points scored by this player. This application can be used by coaches, statistics frameworks, and so on.

We test this application by randomly selecting a player and performing operations. We also test the navigation between activities.

Table 3 shows a comparison of the time taken to perform test cases requiring 10, 25, 50 up to 1 million operations by all the tools. Similar to the first benchmark, Robolectric and MoDroid outperform other tools. Moreover, the time taken by test cases that require one million operations with Robolectric is 118 seconds as opposed to 12 seconds using MoDroid.

## 9 Related Work

### 9.1 Android Mobile Development

This paper advocates the use of modeling to improve the development of Android applications. Modeling pars of an application simplifies and accelerates the development process and frees the developer from writing repetitive code.

The use of models in the development of Java applications has received a lot of attention, and several tools are available. For instance, Eclipse Modeling Framework (EMF) [17] is a powerful modeling tool based on two metamodels Ecore, and Genmodel. EMF stores the model information using XMI (XML Metadata Interchange), and creates its meta-model via UML, Java annotations, XML Schema, and XMI. Similarly, Xcore [6], another tool from Eclipse, is a textual syntax for Ecore. Both EMF and Xcode are powerful tools when it comes to modeling Java applications. However, to the best of our knowledge they have not been used to develop Android applications.

Mobile development frameworks are usually categorized into native, cross-platform, and web based. A native mobile development framework generates applications in native code. Each of those categories has its advantages, and disadvantages. For example, native has the best performance, while web based allows for the fastest development. We compare our approach with some of the frameworks in those categories:

- **native:** App Inventor 2 [12] is a GUI-based tool which supports the rapid development for simple applications. However, when it comes to complex applications, App Inventor 2 sets a lot of limits on the developer, and the application itself since users cannot write their own code, and are only limited to what is provided by the GUI.
- **hybrid:** PhoneGap [18] and Cordova [2] are two commonly used cross platform mobile development frameworks [13]. They allow the developer to generate mobile applications that work on almost all devices by using HTML, CSS, JavaScript. Using JavaScript to interact with the phone's features prevents from using native code since JavaScript is slower in processing data. Moreover, these frameworks lack the ability for background processing, which might be important in several applications. Furthermore, performance issues were reported due to the lack of hardware CSS acceleration of Android [19].

| Operations (#) Platform | 10 | 25 | 50 | 75 | 100 | 150 | 1000 | 10000 | 100000 | 1000000 |
|---|---|---|---|---|---|---|---|---|---|---|
| Robotium | 36 | 88 | 180 | 268 | 360 | 541 | 3605.4 | > 10 hours | > 10 hours | > 10 hours |
| Espresso Emulator | 1.8 | 4.3 | 8.1 | 12 | 15.6 | 23.1 | 159.3 | 1645.5 | 16411.7 | > 10 hours |
| Espresso Sony Z2 | 0.9 | 2.4 | 4.5 | 6.8 | 8.9 | 13.4 | 88.6 | 918.9 | 9189 | > 10 hours |
| Robolectric | 4.4 | 4.5 | 4.7 | 4.9 | 5.1 | 5.4 | 5.6 | 5.9 | 6.9 | 27 |
| MoDroid | 0.021 | 0.031 | 0.038 | 0.039 | 0.04 | 0.05 | 0.14 | 0.4 | 1.118 | 7.7 |

**Table 2.** Testing Time (in seconds).

| Operations (#) Platform | 10 | 25 | 50 | 75 | 100 | 150 | 1000 | 10000 | 100000 | 1000000 |
|---|---|---|---|---|---|---|---|---|---|---|
| Robotium | 5.2 | 12.8 | 25.8 | 37.1 | 49.9 | 73.5 | 486.1 | 4861.1 | > 10 hours | > 10 hours |
| Espresso Emulator | 3.1 | 7.6 | 15.5 | 22.5 | 29.9 | 43.6 | 290.9 | 2845.1 | 28760.2 | > 10 hours |
| Espresso Sony Z2 | 1.1 | 2.6 | 5.5 | 7.7 | 10.3 | 15.3 | 111.9 | 1148.5 | 11275.2 | > 10 hours |
| Robolectric | 4.81 | 4.94 | 5.1 | 5.3 | 5.64 | 5.95 | 6.3 | 8.88 | 18.94 | 118.84 |
| MoDroid | 0.01 | 0.02 | 0.03 | 0.04 | 0.06 | 0.07 | 0.19 | 0.62 | 1.78 | 12.14 |

**Table 3.** Testing Time (in seconds).

- **web-based:** jQuery mobile [10] is one of the most used web based mobile development frameworks. It allows for extremely rapid development of responsive web sites, and applications which can be accessed via all smartphone, tablet, and desktop devices. Two main disadvantages arise when using web based frameworks: poor performance [16], and loosing the ability to use smartphone features.

Finally, none of the above Android development frameworks allows for the composition and decomposition of applications. MoDroid allows for this, as shown in Section 3. Moreover, it allows for permission auto-detection and generation as specified in Section 4. The main advantage is that any unneeded permission will not be included in the Android Manifest file allowing the application to be available for more devices, and most importantly protecting the user's privacy when using additional unneeded permissions [5] [3].

### 9.2 Testing Android Applications

On the other hand, testing of android applications become more challenging. In general, android testing tools can be divided into two main categories: GUI based testing and non-GUI based testing.

***GUI based testing:*** This category requires testing on an emulator or on a real android device. Google present several tools some of which fall under this category. First is Instrumentation [7], a set of classes and methods which control Android components and how Android loads applications. These classes allow the developer to test any component at any given time in its lifecycle. Developing a test case with this tool is time consuming and very complex. This lead Google to develop another tool Espresso [9]. Espresso is built over Instrumentation and its main goal is to simplify testing techniques.

Another commonly used tool is Robotium [15]. This tool is well documented and could be easily configured. In addition to the above, developing test cases is simple; all action calls are being done on a single object `solo`. The main disadvantage one would face using this tool is the speed of running test cases.

Other tools under this category parse applications and automatically generate test cases, e.g., Monkey [8], Android GUITAR [1] and ORBIT [20].

Whether on an emulator or on a real android device, running an enormous number of test cases would require a huge amount of time (see Section 8). This would make GUI based testing tools fall a lot behind non-GUI based testing tools. On the other hand, GUI based testing is more expressive and would be useful to test hardware devises (e.g., camera, sensors, etc.).

***Non-GUI based testing:*** Robolectric [11] allows developers to test Android applications without the use of an Android emulator or device. Robolectric presents the user with several objects and methods to imitate an android application's lifecycle. The main advantage is the speed of running test cases. We would be able to perform thousands of operations by the time GUI based testing is able to perform just tens. Configuring this tool as well as writing test cases are complicated and time consuming. Moreover, it is dependent of several other libraries. In addition, developing test cases is complicated. For instance, Listing 17 is a sample code to access the value of a `TextView` using Robolectric. Our framework falls

**Listing 17.** Sample code to access the value of a `TextView` using Robolectric.

```
1  ActivityClassName activity = Robolectric.
2      buildActivity(ActivityClassName.class).
3      create().start().visible().get();
4
5  TextView results = (TextView) activity2.
6      findViewById( viewID );
7
8  results.getText();
```

under the category of non-GUI based testing. We target ease of configuration, simplicity and performance.

## 10  Conclusion and Future Work

This paper proposes a new way to develop Android applications. It proposes a compromise between expressiveness and ease of development: at the price of slightly reduced expressiveness, MoDroid facilitates and speeds up the development process. Yet, using our framework does not prevent developers from building applications using the full range of features of Android because, after automatically generating the base of the application, expert developers can still use Android features by completing the generated code template. Moreover, our framework introduces several interesting features for developers: decomposition of applications for parallel development, automatic detection of permissions and generation of Manifest, efficient model-based testing of applications, and automatic code generation of some parts of applications.

In the near future, we plan to add several features in the road-map of MoDroid. First, we plan to add emulators for hardware components such as the GPS and camera. For instance, this should allow the user to pre-define GPS locations to be passed to the application. Moreover, we plan to extend MoDroid to support a high-level description of multi-tasking, services, broadcast receivers, etc. Additionally, we plan to make automatic permission detection compatible with the permissions model of the latest version of Android (Android M). Finally, we plan to make MoDroid compatible with existing tools for automatic test generation for Android.